\documentclass[prl,aps,twocolumn,nofootinbib,10pt, showpacs,superscriptaddress]{revtex4-1}
\usepackage{amssymb}
\usepackage{amsmath}
\usepackage{appendix}
\usepackage{times}
\usepackage{graphicx}
\usepackage{color}
\usepackage{lipsum}
\usepackage{hyperref}
\usepackage{nicefrac}
\usepackage{physics}
\usepackage[utf8]{inputenc}
\usepackage{bm}
\usepackage{siunitx}

\newcommand{\dbb}{d_\mathrm{BB}}
\newcommand{\hb}{h_\mathrm{B}}
\newcommand{\hs}{h_\mathrm{S}}
\newcommand{\EF}{E_\mathrm{F}}
\newcommand{\Bem}{B_\mathrm{em}}
\newcommand{\hatn}{\hat{\mathbf{n}}}

\newcommand{\PGI}{Peter Grünberg Institut and Institute for Advanced Simulation, Forschungszentrum Jülich and JARA, 52425 Jülich, Germany}
\newcommand{\RWTH}{Department of Physics, RWTH Aachen University, 52056 Aachen, Germany}
\newcommand{\mainz}{Institute of Physics, Johannes Gutenberg-University Mainz, 55099 Mainz, Germany}
\newcommand{\island}{Science Institute of the University of Iceland, VR-III, 107 Reykjav\'{i}k, Iceland}

\begin{document}

\title {Distinct magnetotransport and orbital fingerprints of chiral bobbers}

\author{M. Redies}
\email{m.redies@fz-juelich.de}
\affiliation{\PGI}
\affiliation{\RWTH}

\author{F. R. Lux}
\affiliation{\PGI}
\affiliation{\RWTH}

\author{J.-P. Hanke}
\affiliation{\mainz}

\author{P. M. Buhl}
\affiliation{\PGI}
\affiliation{\RWTH}

\author{G. P. M\"uller}
\affiliation{\PGI}
\affiliation{\island}

\author{N. S. Kiselev}
\affiliation{\PGI}

\author{S. Bl\"ugel}
\affiliation{\PGI}

\author{Y. Mokrousov}
\email{y.mokrousov@fz-juelich.de}
\affiliation{\PGI}
\affiliation{\mainz}

\begin{abstract}
While chiral magnetic skyrmions have been attracting significant attention in the past years, recently, a new type of a chiral particle emerging in thin films $-$ a chiral bobber $-$ has been theoretically predicted and experimentally observed. 
Here, based on theoretical arguments, we 
uncover that  these novel chiral states possess inherent transport fingerprints that allow for their unambiguous electrical detection in systems comprising several types of chiral states. We reveal that unique transport and orbital characteristics of bobbers root in the non-trivial magnetization distribution in the vicinity of the Bloch points,
and demonstrate that tuning the details of the Bloch point topology can be used to 
drastically alter the emergent response properties of chiral bobbers to external fields, which bears great potential for spintronics applications and cognitive computing.   
 
\end{abstract}

\maketitle
\date{\today}
 Nowadays, chiral magnetic skyrmions are believed to serve as one of the fundamental blocks for future magnetic technologies, such as racetrack memories~\cite{Fert_skyrmion_track} or artificial neurons~\cite{neurons1, neurons2, neurons3}. These fascinating topologically protected chiral particles can be characterized with the quantized flux of the ``emergent" magnetic field $\Bem\sim \hatn \cdot (\partial_x \hatn \times \partial_y \hatn )$ (i.e.~the density of topological charge) due to the spatially non-trivial distribution of spins $\hatn(x,y)$. They also display a number of dynamical effects which make them promising potential bits for efficient creation  and manipulation by external fields~\cite{Nagaosa2013, Fert_skyrmion_track}.  One of the most crucial aspects for the implementation of skyrmionic devices is the ability to distinguish the emergence and dynamics of skyrmions by referring to electronic transport measurements, which are normally associated with the  topological Hall effect arising from the presence of $\Bem$~\cite{Maccariello2018, Franz2014}. 
On the other hand, it was recently predicted theoretically and subsequently confirmed experimentally \cite{Zheng2018}, that in thin films of chiral magnets an intricate interplay of external fields,
temperature and exchange interactions
can result in the formation of novel chiral particles -- chiral bobbers \cite{Rybakov2016}. In contrast to skyrmions that form in tubes~\cite{bobber_paper}, chiral bobbers are localized at the surface and manifestly incorporate a so-called Bloch point (BP) into their structure. These are characterized by fast non-adiabatic changes of the local magnetization around them. 

The experimental discovery of bobbers~\cite{Zheng2018} represents an important milestone in magnetism.  While earlier it was assumed that in chiral magnets there is only one type of particle-like objects $-$ skyrmions $-$ the work by Zheng and co-authors shows that the physics of quasiparticles in chiral magnets is significantly richer, and at least two types of particles with different physical properties may coexist in the same sample. Similar to elementary particles such magnetic quasi-particles may interact with each other with attractive or repelling forces controlled by the strength of the external magnetic field. The discovery of a new type of magnetic quasi-particle opens the vista for new research aiming to investigate the lifetime and mobility of such particles, their collision, and fusion as well as the possible discovery of other, so far unknown, magnetic quasi-particles.
The discovery of bobbers has also possible practical implications, since  the simultaneous presence of skyrmions and bobbers in the same racetrack can significantly simplify the realization of skyrmion-based memories~\cite{Zheng2018}. 
In this context, an important fundamental challenge lies in achieving a clear distinction between various chiral magnetic states, such as skyrmions and bobbers, by most reliable and efficient means of magnetotransport measurements.

In this work, based on microscopic theoretical arguments, we reveal that the two distinct topological phases of skyrmions and chiral bobbers are distinguishable by means of electronic transport. 
We demonstrate that the transition between these topological phases is accompanied by a drastic change in the Hall conductance and an overall enhancement of the orbital moment.
We uncover that this effect is a direct consequence of the complex interplay between the strongly non-collinear distribution of spins on the atomic scale occurring at the tip of the bobber, i.e., at the Bloch point, and effects of electronic hybridization. 
We found that while the Hall and orbital signature of skyrmions is insensitive to the thickness of the sample, the Hall signature of bobbers exhibits a strong linear thickness dependence, which suggests a duality between Bloch points in real space and Weyl points in Weyl semimetals.
We demonstrate that this linear scaling  provides the key to engineering a skyrmion-bobber-racetrack with all-electrical read-out. We also uncover the extreme sensitivity of the transport properties exhibited by the chiral bobbers to their structure, and speculate that this makes them promising candidates as basic elements for reservoir computing~\cite{neurons3, reserv}. This provides a solid foundation for the integration of these  particles into the future generation of memories and devices~\cite{Zheng2018}.


\begin{figure*}
    \centering
    \includegraphics[width=0.91\textwidth]{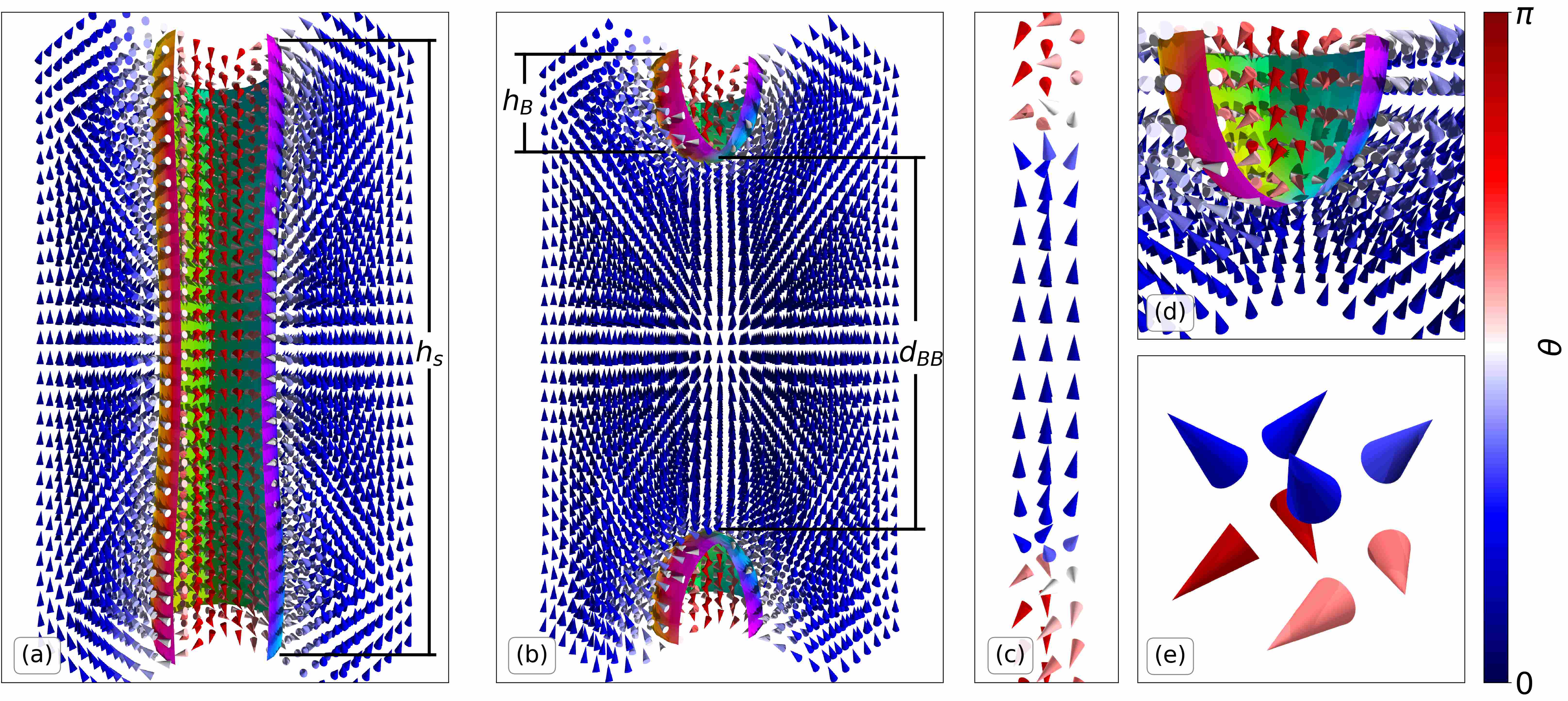}
    \caption{Spin structures of the considered systems. \textbf{(a)} The magnetic structure of the skyrmion tube. The isosurface indicates regions of in-plane magnetization. The color of the spins indicates the polar angle of the magnetization according to the color scale on the right. $\hs$ indicates the slab thickness. \textbf{(b)} 
    The magnetic structure of the twin surface bobber ($h_B=5$). The close-up on the single-surface bobber is shown in (d). $\hb$ stands for the thickness of the bobber, $\dbb$ stands for the distance between the Bloch points of the twin-surface bobber. In \textbf{(c)} the central column of spins of the structure in (b) is shown.
    \textbf{(e)} Eight nearest spins around the Bloch point at the bottom of a single-surface bobber in \textbf{(d)}.
    }
    \label{fig:structures}
\end{figure*}

In this work we focus on thin films, infinite in the $(x,y)$-plane, with a thickness $\hs$  of up to 30 layers in the $z$-direction, see Fig.~\ref{fig:structures}(a). For transport and electronic-structure calculations the skyrmion tubes and bobbers were considered on a 14$\times$14$\times \hs$ cubic grid periodically repeated in-plane, so that a square lattice of spin textures was formed. 
The parametrizations for the skyrmion tube (ST), Fig.~\ref{fig:structures}(a),  \emph{single-surface Bobber} (SSB),  Fig.~\ref{fig:structures}(e), and \emph{twin-surface Bobber} (TSB), Fig.~\ref{fig:structures}(b), were obtained by initializing the spin dynamics code \emph{spirit}~\cite{spirit} with analytic approximations of these structures and relaxing these until convergence was achieved. 
In order to mimic the orbital complexity and electronic properties of thin transition-metal  films, the  electronic structure of the studied spin textures was modelled within the tight-binding approximation for $p$-electrons,
assuming nearest-neighbor hopping on a cubic lattice. The values of the  hopping
integrals ($V_{pp\sigma} = \SI{-1.5}{\electronvolt}$, $V_{pp\pi} = \SI{-0.3}{\electronvolt}$), exchange splitting of 
$\SI{1.4}{\electronvolt}$ and spin-orbit strength of $\SI{0.4}{\electronvolt}$ 
correspond to typical values in thin film and multilayered transition-metal systems which exhibit skyrmionic states~\cite{typical_param1, typical_param2}.
To access the Hall transport properties~\cite{Nagaosa2015,Mertig2018}  we used the Berry curvature expression for the intrinsic Hall conductance, where the Berry curvature is
$\Omega_{xy}^n (\bm{k}) = -2\Im \braket{\partial_{k_x} u_{n\bm{k}}}{\partial_{k_y} u_{n\bm{k}}}$, with $u_{n\bm{k}}$ as the 
lattice-periodic part of the wavefunction of band $n$ at Bloch vector $\mathbf{k}$. Using this expression, we arrive at the values for the intrinsic Hall conductance per layer, which we refer to as the Hall conductivity (HC) in the following. It is given by $\sigma_{xy} = \frac{1}{\hs}\sum_n^{\mathrm{occ}} \int_{\mathrm{BZ}} \Omega_{xy}^n(\bm{k})  \, d \bm{k}$ (in units of $e^2/h$).
To access the orbital magnetization of the textures we used the formalism of the modern theory~\cite{thonhauser_om,thonhauser_om2, xiao_om} to arrive at the values for the orbital moment per unit cell, $\bm{M}$. For further details see Supplementary material~\cite{Supp}.

\begin{figure}
    \includegraphics[width=0.5\textwidth]{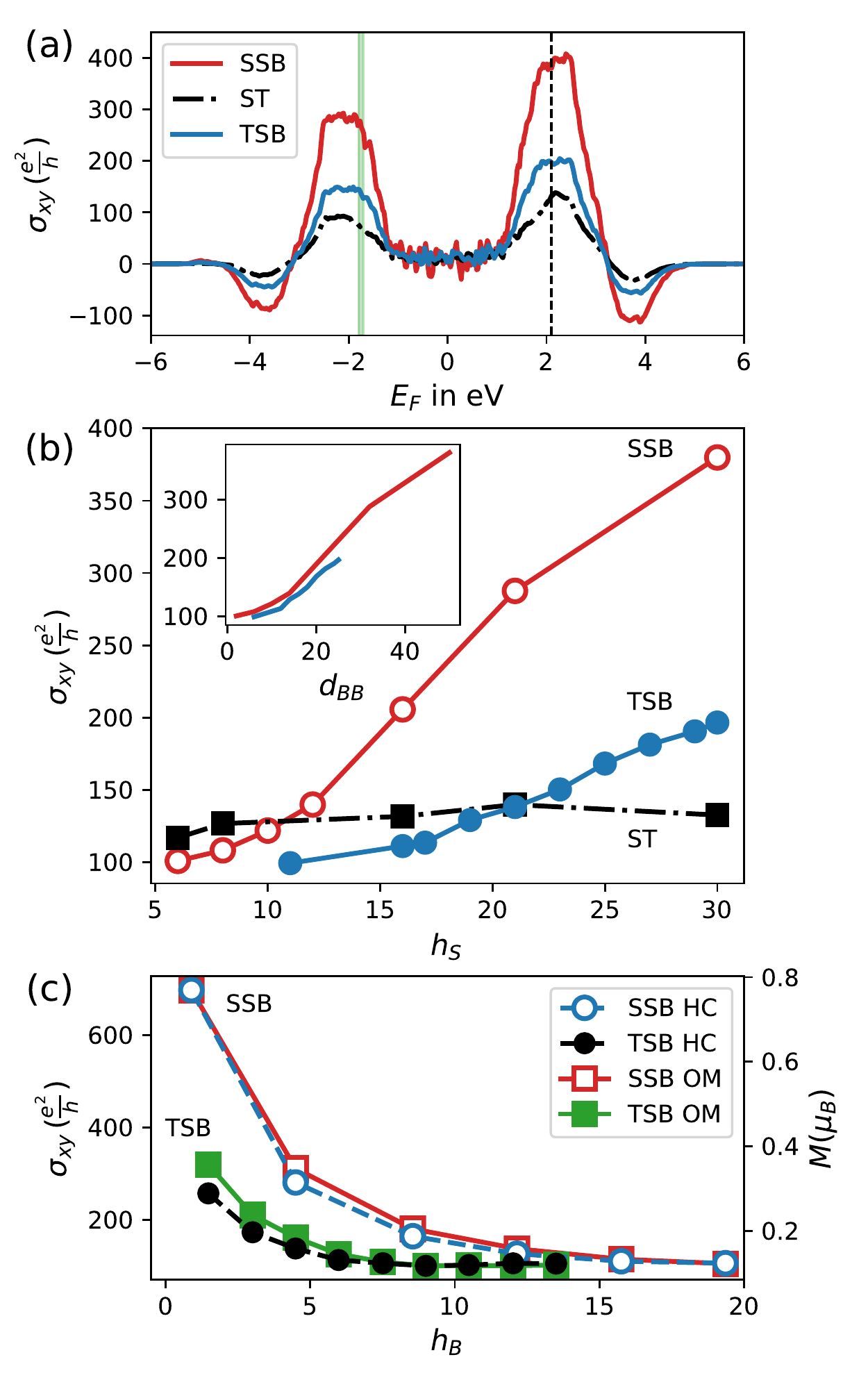}
    \caption{Hall transport properties of chiral bobbers versus skyrmion tubes. 
    \textbf{(a)} Hall conductivity of the ST (black), SSB (green), TSB (blue)
    is shown for $\hs=30$ and $\hb=5$ as a function of the Fermi energy $\EF$. 
    \textbf{(b)} Hall conductivity at $\EF = 2.1$\,eV, marked with a thin dashed line in (a), is plotted as a function of the slab thickness $\hs$. 
    The color code is the same as in (a). 
    The inset shows $\sigma_{xy}$ of SSB and TSB textures as a  function of the distance between the Bloch points.
    \textbf{(c)} $\sigma_{xy}$ is shown as a function of the height $\hb$ of the SSB (open blue circles) and TSB (solid black circles).  $M$ is indicated by the axis on the right as a function of $\hb$ for the SSB (open red squares) and TSB (solid green squares). For very large values of $\hb$ both the SSB and TSB will transform into ST-like systems, thus converging to a common value of the HC.}
    \label{fig:lin}
\end{figure}

%
%

We start by analysing the Hall transport properties of the skyrmion tube texture, see  Fig.~\ref{fig:structures}(a). The Hall conductivity of a ST as a function of the Fermi energy $E_F$, presented in Fig.~\ref{fig:lin}(a) for $\hs=30$, exhibits a double-peak structure  resembling the total electronic density of states (DOS) in the system~\cite{Supp}. 
By varying the thickness of the film 
we discover that the HC of the ST varies only slightly, as exemplified  in Fig.~\ref{fig:lin}(b) for $\EF=\SI{2.1}{\electronvolt}$. The corresponding topological robustness of the Hall transport is further confirmed by the small variation of $\sigma_{xy}$ in response to in-plane shrinking or expanding of the ST texture~\cite{Note1}. While such behavior could be expected from the topological Hall effect of skyrmions in the adiabatic limit of a very large texture without spin-orbit interaction~\cite{Bruno2004,nagaosa_tokura}, 
it is remarkable to find a similar stability for orbitally complex  strongly spin-orbit coupled states. Here one might expect an ultra-sensitive dependence on deformations of the magnetic structure which is not observed in our data. Our findings firmly put the ST type of the spin texture into a distinct topological class which can be characterized by a robust value of the intrinsic $\sigma_{xy}$, the exact value of which is determined by the electronic-structure details.

On the other hand, the behavior of $\sigma_{xy}$ of chiral bobbers is manifestly different. While the overall dependence of the HC on $E_F$ is qualitatively very similar for the ST, SSB and TSB textures, see Fig.~\ref{fig:lin}(a), and the total DOS of all three systems are basically indistinguishable~\cite{Supp}, both types of bobbers exhibit linear growth in $\sigma_{xy}$ as a function of the film thickness, Fig.~\ref{fig:lin}(b). As a result, the HC of the single surface bobber significantly exceeds that of the ST over a large range of $h_S$, as exemplified for $\EF=2.1$\,eV in Fig.~\ref{fig:lin}(b), being as much as three times larger than the ST $\sigma_{xy}$ for $\hs=30$. 
Notably, this result, both in terms of general trend as well as absolute magnitude of the Hall conductivities, pertains over a wide region of spin-orbit strength down to 0.1\,eV~\cite{Supp}.   We believe that this finding motivates the design of a system, where various textures
can be distinguished by magnetotransport measurements only.
Since according to our calculations the HC of ferromagnetic films is negligible irrespective of $h_S$,
the distinct Hall transport properties of bobbers must originate in
the strong variation of the  magnetization in the vicinity of the BPs.

The characteristic distribution of the spins around the BPs in  bobbers, depicted in Fig.~\ref{fig:structures}(d,e), reflects the fact that  they present a special type of effective ``monopoles" emerging in real-space~\cite{monopole1, monopole2}. This is best understood taking the TSB as an example: while the spins around the upper Bloch point in Fig.~\ref{fig:structures}(c) all point towards it, the lower BP has all surrounding spins pointing away from it, and thus the lower and upper BPs act as a source and sink of the magnetization, respectively. In fact, a certain resemblance exists between the properties of the TSB (with Bloch points along $z$ separated by $d_{BB}$) and a magnetic  Weyl semimetal~\cite{Weyl1,Weyl2} exhibiting a pair of Weyl points of opposite chirality in reciprocal $\mathbf{k}$-space (e.g. along $z$ separated by $k_w$), which serve as sources and sinks of $\mathbf{k}$-space Berry curvature. Notably, in case of an ideal Weyl semimetal, it can be shown that the intrinsic anomalous HC obeys the law of $\sigma_{xy}\sim k_w$~\cite{Burkov2011}, which is in a qualitative agreement to the behavior of the 
HC of a TSB system when replacing the $k$-space distance $k_w$ with the distance between the
BPs in real space, $d_{BB}$, see inset of Fig.~\ref{fig:lin}(b). Moreover, the difference in the Hall properties between the SSB and TSB textures can be grasped 
after introducing an ``image charge"~\cite{monopole1,monopole2}  of the SSB slab by adding its mirror image with respect to the lower surface of the film, and thereby defining $\dbb$ as the distance between the original single Bloch point and its image of opposite charge. Thus, in the case of a SSB, $\dbb$ corresponds to twice the distance between the Bloch point and the bottom surface of the slab. Introducing  this unified $\dbb$ parameter reveals a remarkable agreement between the HC curves for the SSB and TSB,
see inset in Fig.~\ref{fig:lin}(b).

One of the main differences between the ST and bobber textures is the spatial localization and magnitude of the local emergent magnetic field $\Bem$ which is qualitatively proportional to the local scalar chirality between neighboring spins,   $\Bem\sim \mathbf{S}_i\cdot(\mathbf{S}_j\times \mathbf{S}_k)$~\cite{libor, Nagaosa2010, nagaosa_tokura}. 
While the overall flux of $\Bem$ through any layer of the ST and bobber textures is similar, the local magnitude of $\Bem$ at the BP can exceed the averaged $\Bem$ in the ST by an order of magnitude.
Here, the adiabatic assumption which is the defining idea behind the emergence of $\Bem$ cannot be expected to be valid.
This has a drastic effect on the properties of the electronic states emerging in bobber textures as discussed in the following. It also leads to a pronounced dependence of the HC on the details of the distribution $\Bem$ around the Bloch points. This stands in sharp contrast to the ``topological" behavior of the Hall effect in STs or Weyl semimetals. To convince ourselves of the latter, we monitor the changes in the HC as we scale the shape of the SSB along $z$ and change its height $\hb$, see Fig.~\ref{fig:structures}(b), thereby tuning the spatial extent and local magnitude of $\Bem$. As a result of this, we observe a drastic enhancement of the HC with decreasing $\hb$ for SSBs and TSBs, see Fig.~\ref{fig:lin}(c). 

\begin{figure}
    \includegraphics[width=0.46\textwidth]{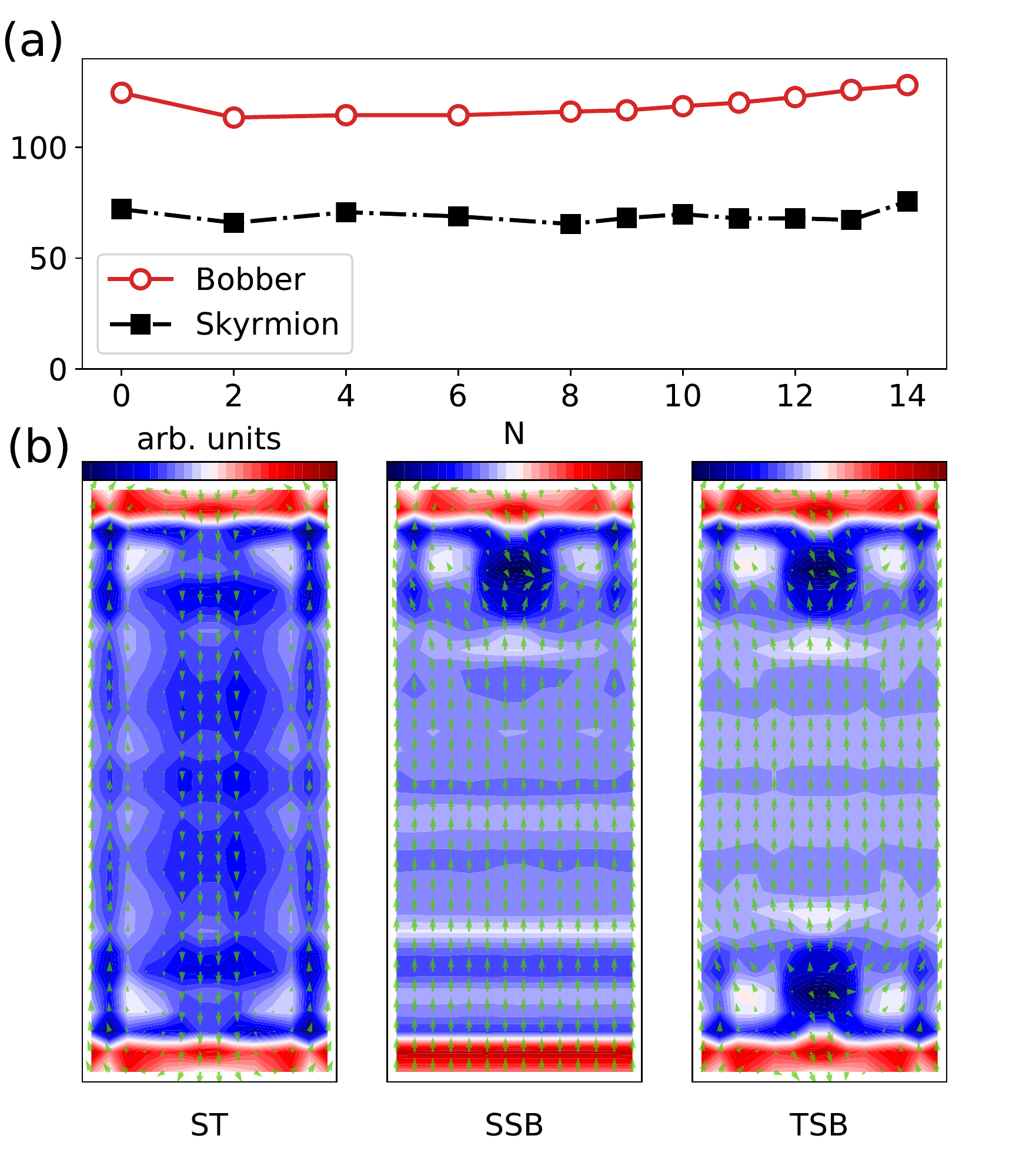}
    \caption{\textbf{(a)} The magnitude of the Hall conductance of a 3-layer slice of the 17-layer thick film containing a ST, and a SSB ($\hb=5$), as a function of the position of the slice in the film (see text for details).
    \textbf{(b)} A cut of the local DOS  through the middle of the unit cell integrated from $-1.8$\,eV to $-1.7$\,eV (indicated in green in Fig.~\ref{fig:lin}a) for a ST, SSB and TSB with $\hs=30$ and $\hb=5$.  }
    \label{fig:state_analysis}
\end{figure}

On the other hand, remarkably, the pronounced transport of bobbers  is not dominated by the surface contribution, as one might anticipate given their real-space confinement.
 We show this by computing the layer-resolved HC, which 
characterizes the magnitude of the Hall current flowing within a thin slab 
of thickness $L$ between layers $N$ and $N+L$ of the film in response to the
electric field applied within the same region (for more details see Supplementary~\cite{Supp}).
We can conclude from our calculations for a SSB and a ST performed 
for $L=3$, $h_S=17$ and $E_F=2.1$\,eV and presented in Fig.~\ref{fig:state_analysis}(a), that the overall rise of the HC in the bobber system as compared to the ST in this case can be attributed to the enhancement of the Hall current
more or less uniformly within the film.
This points at the fact that the presence of the Bloch points results in a formation of delocalized electronic states (via imposing the boundary conditions on their spin part) which carry large Berry curvature. An example of such states can be seen in Fig.~\ref{fig:state_analysis}(b). 
In this plot, the apparent non-trivial behavior of the states across the film in bobber systems stands in sharp contrast to the relatively uniform in $z$ nature of the electronic states in STs. The linear behavior of the $\sigma_{xy}$ of bobbers in Fig.~\ref{fig:lin}(b) can be then explained by the thickness-dependent enhancement of the geometrical ``strength" (as given by the Berry curvature) of the corresponding emergent states whose properties are also 
extremely sensitive to the boundary conditions encoded in the details of the spin distribution around the Bloch points.

The findings of our paper concerning transport properties are applicable to the orbital magnetism exhibited by the textures as well. In recent years, it was shown from model considerations, tight-binding and {\it ab-initio} studies, that the presence of non-zero scalar spin chirality in large skyrmions as well as in strongly frustrated spin systems is inevitably associated with pronounced contributions to the orbital magnetization (OM), which arises in response to the emergent field $\Bem$~\cite{Hoffmann2015, dosSantosDias2016, Hanke2016, Hanke2017, fabians_paper}. Indeed, our calculations indicate the non-vanishing topological contribution to the OM in STs, however, an intriguing finding is that the OM of chiral bobber systems is very closely correlated with the behavior of the HC, see for example Fig.~\ref{fig:lin}(c). Given the enhancement of the Hall signal of TSBs and especially SSBs, as compared to STs, we speculate that the corresponding enhancement of the orbital magnetism could be used to detect and distinguish different chiral textures by referring to orbitally sensitive techniques such as XMCD~\cite{fabians_paper}. 

The pronouncedly much  stronger sensitivity of the transport characteristics of chiral bobbers versus skyrmion tubes to the fine details of the spin distribution marks them as 
very attractive candidates for reservoir computing~\cite{neurons3, reserv}. Indeed, the non-linear voltage characteristics exhibited by pinned skyrmions have been pivotal for suggesting skyrmions as basic elements 
for this particular realization of cognitive computing~\cite{neurons3}. This is based on the observation of the dependence of skyrmion's resisitivity on texture deformations brought
by the applied current $-$ the effect which should be much more prominent in chiral bobbers.
In addition to exploiting the current-induced spin torques for modifying the shape of the bobber, we suggest that the recently observed effect of current-induced Dzyaloshinskii-Moriya interaction (DMI) could be also used for this purpose~\cite{karnad,hayashi}.
Within the latter scenario, the applied current can modify the strength of the DMI in the system in a manifestly non-linear fashion with respect to the current density~\cite{karnad,hayashi}. Given the direct dependence of the bobber depth $h_B$ on the DMI strength~\cite{bobber_paper}, this should in turn result in strongly non-linear current-voltage characteristics of chiral bobbers.

To conclude, the central result of our paper, that is  the linear scaling of the Hall signal with the bobber separation,  
opens a way to engineering the transport properties by thickness and electronic structure design in such a way that chiral bobbers can be distinguished from skyrmion tubes by all-electrical transport measurements. This is a highly  promising opportunity which can be exploited in future racetrack devices. 
Moreover, we speculate that the pronounced difference in Hall transport  of various textures that we observe inevitably extends to longitudinal transport and tunneling  properties, such as~e.g.~magnetoresistance, anisotropic magnetoresistance and scanning tunneling microscopy fingerprints, as well as to current-driven dynamical phenomena such as the bobber Hall effect and current-driven bobber motion. 
Overall, our findings open up a whole new horizon of possibilities associated with the diversity of emergent chiral dynamics and chiral transport, and gear us toward an integration of the zoo of topological chiral phases in future technologies.

\begin{acknowledgments}
We gratefully acknowledge computing time on the supercomputers JUQUEEN and JURECA at J\"ulich Super-computing Center,  and at the JARA-HPC cluster of RWTH Aachen. We  acknowledge  funding  under SPP 2137 ``Skyrmionics"   and project  MO  1731/5-1  of  Deutsche  Forschungsgemeinschaft (DFG). This work has been also supported by the DFG through the Collaborative Research Center SFB 1238, as well as by the DARPA
TEE program through grant MIPR\# HR0011831554 from DOI.
\end{acknowledgments}

\bibliography{mainBib}

\end{document}